\documentclass[sigconf]{acmart}
\usepackage{multirow}
\usepackage{amsfonts}
\usepackage[caption=false]{subfig}
\usepackage{textcomp}
\usepackage{stfloats}
\usepackage{url}
\usepackage{enumitem}
\usepackage{balance}
\usepackage[linesnumbered,ruled,vlined]{algorithm2e}
\AtBeginDocument{%
  }

\setcopyright{cc}
\setcctype{by-nc-nd}
\copyrightyear{2026}
\acmYear{2026}
\acmDOI{10.1145/3770854.3780275}
\acmConference[KDD '26]{Proceedings of the 32nd ACM SIGKDD Conference on Knowledge Discovery and Data Mining V.1}{August 09--13, 2026}{Jeju Island, Republic of Korea}
\acmBooktitle{Proceedings of the 32nd ACM SIGKDD Conference on Knowledge Discovery and Data Mining V.1 (KDD '26), August 09--13, 2026, Jeju Island, Republic of Korea}
\acmISBN{979-8-4007-2258-5/2026/08}
\acmPrice{}
\begin{document}

\title{\textsc{APT-CGLP}: Advanced Persistent Threat Hunting via Contrastive Graph-Language Pre-Training}

\author{Xuebo Qiu}
\affiliation{%
	\institution{College of Computer Science and Technology, Zhejiang University of Technology}
	\city{Hangzhou}
	\country{China}
}
\email{xueboqiu@zjut.edu.cn}

\author{Mingqi Lv}
\affiliation{%
	\institution{College of Geoinformatics, Zhejiang University of Technology}
	\institution{Zhejiang Key Laboratory of Visual Information Intelligent Processing}
	\city{Hangzhou}
	\country{China}
}
\email{mingqilv@zjut.edu.cn}
\authornote{Corresponding author.}

\author{Yimei Zhang}
\affiliation{%
	\institution{College of Computer Science and Technology, Zhejiang University of Technology}
	\city{Hangzhou}
	\country{China}
}
\email{yimeizhang0229@zjut.edu.cn}

\author{Tieming Chen}
\affiliation{%
	\institution{College of Geoinformatics, Zhejiang University of Technology}
	\institution{Zhejiang Key Laboratory of Visual Information Intelligent Processing}
	\city{Hangzhou}
	\country{China}
}
\email{tmchen@zjut.edu.cn}

\author{Tiantian Zhu}
\affiliation{%
	\institution{College of Computer Science and Technology, Zhejiang University of Technology}
	\city{Hangzhou}
	\country{China}
}
\email{ttzhu@zjut.edu.cn}

\author{Qijie Song}
\affiliation{%
	\institution{College of Geoinformatics, Zhejiang University of Technology}
	\city{Hangzhou}
	\country{China}
}
\email{songqijie@zjut.edu.cn}

\author{Shouling Ji}
\affiliation{%
	\institution{ College of Computer Science and Technology,
		Zhejiang University}
	\city{Hangzhou}
	\country{China}
}
\email{sji@zju.edu.cn}

\renewcommand{\shortauthors}{Qiu et al.}

\begin{abstract}
  Provenance-based threat hunting identifies Advanced Persistent Threats (APTs) on endpoints by correlating attack patterns described in Cyber Threat Intelligence (CTI) with provenance graphs derived from system audit logs. A fundamental challenge in this paradigm lies in the \textit{modality gap}—the structural and semantic disconnect between provenance graphs and CTI reports. Prior work addresses this by framing threat hunting as a graph matching task: 1) extracting attack graphs from CTI reports, and 2) aligning them with provenance graphs. However, this pipeline incurs severe \textit{information loss} during graph extraction and demands intensive manual curation, undermining scalability and effectiveness.

  In this paper, we present \textbf{\textsc{APT-CGLP}}, a novel cross-modal \underline{APT} hunting system via \underline{C}ontrastive \underline{G}raph-\underline{L}anguage \underline{P}re-training, facilitating end-to-end semantic matching between provenance graphs and CTI reports without human intervention.
  First, empowered by the Large Language Model (LLM), \textsc{APT-CGLP} mitigates data scarcity by synthesizing high-fidelity provenance graph-CTI report pairs, while simultaneously distilling actionable insights from noisy web-sourced CTIs to improve their operational utility. 
  Second, \textsc{APT-CGLP} incorporates a tailored multi-objective training algorithm that synergizes contrastive learning with inter-modal masked modeling, promoting cross-modal attack semantic alignment at both coarse- and fine-grained levels. 
  Extensive experiments on four real-world APT datasets demonstrate that \textsc{APT-CGLP} consistently  outperforms state-of-the-art threat hunting baselines in terms of accuracy and efficiency. 
\end{abstract}



%% information and tument, and typically spans the

%%
%% This command processes theaffiliation and title
\maketitle

\section{Introduction}
Advanced Persistent Threats (APTs) pose a significant risk due to their persistence, stealth, and potentially severe impact \cite{linkedin}. 
To combat these threats, security practitioners engage in the collection, analysis, and interpretation of Cyber Threat Intelligence (CTI) that offers actionable insights into offensive tactics, techniques, and procedures (TTPs) \cite{cti_survey}.
Following this foundation, \textit{threat hunting} has emerged as a proactive defense strategy that utilizes CTI-recorded attack knowledge to identify system compromises \cite{sans}.

Given the complexity of APT attacks, recent studies \cite{provgsearcher,megrapt,flash,kairos} adopt \textit{data provenance} to convert audit logs into provenance graphs (see Figure \ref{motivation}c) that capture structured interactions (e.g., file writing) among system entities (i.e., process, file, socket), enabling comprehensive system behavior analysis.
Capitalizing on this, threat hunting is typically formulated as two categories of matching tasks. The first, \textit{signature-based matching}, searches for CTI-documented Indicators of Compromise (IoCs) within provenance graphs. While efficient, it is highly vulnerable to evasion techniques like IoC obfuscation \cite{pyramidofpain}.
The second, \textit{behavior pattern-based matching}, aims to mine abstract attack patterns from CTI reports and match them with provenance graphs \cite{provgsearcher,ghunter,extractor}.
This approach offers greater generalization and has seen broad adoption in practice \cite{poirot,megrapt,attackg,trec}. Concretely, it adheres to a two-stage process:
1) \textbf{Query Graph Construction}, which derives attack (query) graphs from CTI reports—either manually or via automated extraction framework—to bridge the natural language modality of CTI reports with the graph modality of provenance data; and 
2) \textbf{Graph Pattern Alignment}, which applies heuristic rules or graph learning algorithms to match query graphs against suspicious provenance graphs, aiming to uncover behavior patterns indicative of APT attacks.
Despite their potential, they suffer from a major limitation:
\textit{transforming unstructured CTI reports into query graphs incurs substantial information loss} \cite{megrapt,poirot,provgsearcher}, primarily due to the complexity and ambiguity of attack narratives (see Section~\ref{sec:motivation}). As a result, extensive human intervention is required to recover missing semantics from the query graph, which undermines the efficacy and scalability of such methods.

Meanwhile, multimodal learning has made remarkable strides across various domains \cite{clip,albef,blip,graphtranslator,taga}, with growing emphasis on integrating information from heterogeneous sources to improve cross-modal task performance.
A prominent milestone, CLIP \cite{clip}, attains strong zero-shot capabilities on multiple benchmarks by learning unified representations from large-scale image-caption pairs.
Such advances inspire a pivotal question: \textit{can APT hunting be transformed into a similar end-to-end matching process between provenance graphs and CTI reports that share consistent attack semantics, thereby circumventing the reliance on manual CTI engineering?}
However, unlike image-caption alignment, bridging gaps between provenance graphs and CTI reports introduces unique challenges:
%Building on the success in the vision-language domain, recent research has expanded to joint training of graph and text encoders in a cross-modal contrastive learning manner, aiming to enhance the performance of downstream tasks \cite{g2p2,graphtranslator,taga,grenade,graphgpt}.
%For instance, G2P2 \cite{g2p2} seeks to pre-train a graph-text model for zero-shot low-resource text classification.
%Despite their success, they typically operate in simplified scenarios, such as graphs with isomorphic structures or few types of associations (e.g., citation networks), and text data that is clean and well-defined, and offers strong supervision for learning representations across modalities.
%In contrast, the complexity of threat hunting—where CTI reports are noisy and provenance graphs feature intricate relations, heterogeneous and time-series structures—renders existing graph-text alignment methods ill-suited for our purposes.  The key challenges can be summarized as follows:
  
\textbf{C1. Significant Modality Gaps.} Provenance graphs provide low-level, structured representations of system behaviors from audit logs, while CTI reports contain high-level APT narratives in unstructured natural language. Aligning these disparate semantic abstraction and structural formats remains a major challenge.
  
\textbf{C2. Lack of Cross-Modal Supervision.} Although CTI reports are widely available, matched provenance graphs remain extremely scarce due to the difficulty of acquiring authentic APT data. Meanwhile, the prohibitive cost of APT simulations (e.g., the DARPA TC program \cite{optcanalysis}) further limits the manual curation of training samples, forming a critical bottleneck for multimodal learning.
  
\textbf{C3. Noisy and Verbose CTIs.} Unlike the concise and semantically focused image captions in vision-language tasks, CTI reports are often verbose (averaging over 3,000 words \cite{ctis}) and laden with noisy content (e.g., advertisements), which impedes the effective exploitation of inherent attack knowledge for threat hunting.

To address these challenges, we present \textbf{\textsc{APT-CGLP}}, a pioneering cross-modal \underline{APT} hunting system via \underline{C}ontrastive \underline{G}raph-\underline{L}anguage \underline{P}re-training, enabling end-to-end threat hunting without manual intervention.
For challenge C1, we design a multi-objective training algorithm that integrates contrastive learning and inter-modal masked modeling to comprehensively align attack semantics between provenance graphs and CTI reports. 
%An enhanced training algorithm integrating contrastive learning and cross-modal masked modeling is then designed to align attack semantics at both global and fine-grained levels. 
More specifically, contrastive learning first aligns their global semantic representations, while a cross-attention mechanism operates over their node- and token-level embeddings in conjunction with masked modeling, thereby enforcing fine-grained semantic consistency.
%\textsc{APT-CGLP} incorporates a graph encoder designed for provenance graph and a decoupled transformer-based encoder, where the first half layers encode CTI reports, and the latter half layers serve as a multimodal encoder performing cross-attention with the token-level graph representations to model nuanced feature interactions. 
For challenge C2, we propose a Graph2CTI module that leverages Large Language Models (LLMs) to automatically synthesize CTI-style reports from attack-free provenance graphs to augment training data.
%This module utilizes the In-Context Learning capabilities of LLMs, using carefully curated contextual examples to generate expert-like activity reports based on structured interaction data from provenance graphs.
For challenge C3, we introduce a CTI denoising module powered by the chain-of-thought reasoning capacity of LLMs to distill concise and actionable threat insights from noisy CTI reports.
%By following meticulously designed instructions, LLMs extract focused attack knowledge from noisy CTI reports, ensuring high-quality textual behavioral representations.
%Finally, we design a two-stage threat hunting strategy that frames the process as a retrieval-based task, which involves first retrieving relevant reports from the knowledge base, followed by a trained multimodal encoder for refined selection, ensuring efficient and precise identification of potential threats.
%This method leverages LLaMA 's instruction-following capabilities to produce high-quality summaries of attack descriptions within CTIs, thereby improving the reliability and precision of threat hunting with coherent insights.
Extensive experiments on four real-world APT datasets \cite{darpatce3,darpatcoptc} demonstrate that \textsc{APT-CGLP} outperforms state-of-the-art threat hunting baselines, while eliminating the need for manual query graph engineering.

In summary, the contributions of this work are as follows:
\begin{itemize}[leftmargin=*]
  \item We propose \textsc{APT-CGLP}, an end-to-end APT hunting system that facilitates threat hunting from provenance graphs and CTI reports without requiring human intervention.  %\item We design a Graph2CTI module that harnesses the few-shot learning capabilities of LLMs to generate high-quality paired provenance graph-CTI report samples for training augmentation.
  \item We design a multi-objective training algorithm that combines contrastive learning with inter-modal masked modeling to enhance semantic alignment between provenance graphs and CTI reports at both coarse- and fine-grained levels.
  \item We design a series of LLM-driven optimization strategies to improve generalization, including a Graph2CTI module for synthesizing provenance graph-CTI report training pairs, and a CTI denoising module for streamlining real-world CTI reports.
  %\item We leverage the instruction-following abilities of LLMs with tailored chain-of-thought prompts to distill attack knowledge from CTI reports. 
  \item We conduct extensive experiments on real-world APT attack datasets, demonstrating that \textsc{APT-CGLP} outperforms state-of-the-art methods in both accuracy and efficiency.
\end{itemize}
%-------------------------------------------------------------------------------
\section{Related Work} \label{sec:motivation}
\noindent\textbf{Provenance-based Threat Hunting.}
Current approaches \cite{provgsearcher, megrapt, poirot, deephunter} commonly formulate threat hunting as a graph pattern matching task. They first transform CTI-recorded attack patterns into structured query graphs, and then apply pattern matching algorithms to align them with provenance graphs for APT hunting.

For instance, Poirot \cite{poirot} converts IoC interactions into query graphs, and matches against provenance graphs using a cost-aware similarity function that prioritizes causal relationships and information flows.
Threatraptor \cite{threatraptor} introduces a custom natural language processing (NLP) pipeline for CTI knowledge extraction and a domain-specific threat query language for large-scale audit log analysis. 
Graph learning-based methods like MEGR-APT \cite{megrapt}, DeepHunter \cite{deephunter} extract query graphs through manual CTI analysis, and convert the matching problem into vector similarity evaluation through neural representations. ProvG-Searcher \cite{provgsearcher} utilizes the order embedding technique to learn subgraph entailment relationships, enabling efficient subgraph matching via online comparisons using precomputed provenance graph representations.
\begin{figure}[]
  \centering
  \includegraphics[width=3.3in]{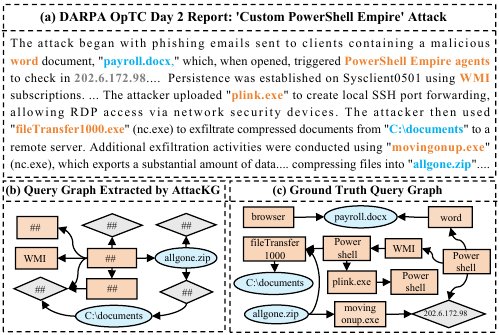}
  \caption{\small{A motivating example, where subfigure (a) denotes the denoised CTI report from the DARPA OpTC engagement, and subfigures (b) and (c) represent the query graphs generated by AttacKG (\#\# denotes unknown entities) and the ground truth, respectively.}}
  \label{motivation}
\end{figure}

Despite their successes, these methods are fundamentally constrained by their \textit{heavy reliance on labor-intensive query graph construction}.
Although NLP techniques have been introduced to automate this transformation, they remain error-prone. Consider a denoised CTI report in Figure \ref{motivation}a, which describes an APT campaign from the OpTC engagement \cite{darpatcoptc} involving PowerShell Empire for data exfiltration. Nonetheless, even after applying our CTI denoising module (see $\S$\ref{sec:cti-denoising}) to refine the report and improve its downstream usability, AttacKG \cite{attackg}, a state-of-the-art CTI extraction framework, still produces a query graph (Figure \ref{motivation}b) that captures only 30\% of the key system entities compared to the ground truth (Figure \ref{motivation}c), severely limiting its effectiveness for threat hunting.
%Critical entities such as payloads (\texttt{payroll.docx}) and their dependencies are entirely missed, undermining threat hunting efficacy.
% of existing two-stage threat hunting systems. , thereby undermining the effectiveness of subsequent graph alignment and threat detection necessitating further labor-intensive manual corrections to recover details.

\noindent \textbf{Multimodal Learning.}
Multimodal learning aims to build models that can jointly process and align information from heterogeneous modalities by learning a shared embedding space \cite{multimodal_survey, zhang2025improving}.
CLIP \cite{clip} is a representative work that aligns visual and textual modalities using a simple yet effective contrastive learning approach, achieving strong zero-shot performance across various tasks.
Follow-up work has extended this paradigm with more sophisticated alignment mechanisms \cite{albef,coca,blip,blip2,cogvlm}. For example, BLIP \cite{blip} introduces a bootstrapping approach that combines contrastive learning with captioning and filtering objectives to better handle noisy web data, while BLIP-2 \cite{blip2} further improves efficiency by incorporating a lightweight querying transformer in a two-stage training framework.
%CogVLM \cite{cogvlm} addresses shallow alignment limitations by integrating trainable visual expert modules, enabling deeper vision-language fusion without degrading NLP capabilities.
Beyond vision-language domains, recent studies have also explored fusing molecular graphs with textual descriptions for molecular retrieval \cite{molecular1,molecular2,molecular3}. 
Nonetheless, multimodal learning remains underexplored in the context of threat hunting, primarily due to the scarcity of training data and the alignment challenges between provenance graphs and CTI reports, as discussed earlier.
%To address these challenges, we propose an optimized data generation module to generate high-quality training data, coupled with an enhanced training algorithm to augment training effectiveness. 
%This approach enables the model to capture both global and fine-grained semantic correlations between provenance graphs and CTIs, bridging the modality gap and delivering a robust, efficient solution for threat hunting.

\section{Preliminaries} \label{prov_def} 
\subsection{Definitions}
\noindent \textbf{Provenance Graph.}
A provenance graph provides a structured representation of audit logs, enabling comprehensive analysis of system behaviors. Formally, it is defined as $G = (V, E)$, where $V$ represents the set of nodes corresponding to system entities (i.e., process, socket, file), and $E$ is the set of directed edges indicating system events. 
Each edge $e \in E$ is defined as the tuple $e = \langle s, a, o, t \rangle $, where $s$ is a subject entity (a process) initiating the action $a$ (e.g., write), $o \in V$ is the object entity, and $t$ is the timestamp of the event.
%One of its subgraphs or paths can represent a specific activity (see Figure \ref{motivation}).

\noindent \textbf{Cyber Threat Intelligence (CTI).}
CTI refers to evidence-based security knowledge, including IoCs, TTPs, and threat assessments (Figure \ref{motivation}a), which supports informed defense decisions \cite{cti_survey}.
CTI reports consolidate threat data from various sources (e.g., audit logs, vulnerability databases, threat intelligence feeds) and are further enriched through manual examination and expert interpretation. Such insights empower organizations to proactively anticipate, identify, and respond to cyber threats more effectively.

\begin{comment}
\noindent \textbf{Problem and Goal.}
The provenance-based threat hunting problem involves proactively identifying provenance graphs that match attacks recorded in CTI reports \cite{megrapt,provgsearcher}. 
Our goal is to develop a model capable of directly aligning the attack patterns of provenance graphs and CTI reports, enabling end-to-end APT hunting without the need for modality transformation or manual intervention.
\end{comment}

%\noindent \textbf{Threat Model}
\subsection{Threat Model}
Our threat model assumes a trusted computing base encompassing the operating system, hardware, and audit frameworks, in line with prior work \cite{kairos,magic,megrapt,aptshield,aptkgl}. Potential threats such as audit log tampering or hardware-level exploits are deemed beyond the scope of this work.
Furthermore, the CTI reports utilized for threat hunting are regarded to be accurate, trustworthy, and free from adversarial manipulation.
While adversaries may adapt the implementation details of attack techniques to evade detection, we posit that the core semantics and objectives of these attacks remain unchanged.

\section{System Architecture}
\begin{figure}[]
  \centering
  \includegraphics[width=3.4in]{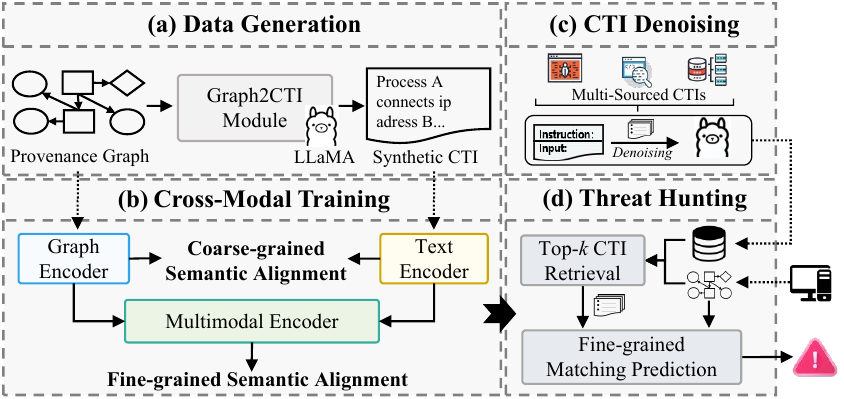}
  \caption{\small{\textsc{APT-CGLP} architecture. (a) Synthetic provenance graph-CTI report pair generation for training; (b) Cross-modal semantic alignment via multi-objective pre-training; (c) CTI report denoising to enhance the usability; and (d) Two-stage retrieval using trained encoders in (b) to balance efficiency and precision in threat hunting.}}
  \label{framework}
\end{figure}

%As illustrated in Figure \ref{framework}, \textsc{APT-CGLP} comprises four modules: 1) a \textbf{Graph2CTI module} that generates paired provenance graph-CTI report datasets using LLMs' in-context learning capabilities, 2) a \textbf{cross-modal training module} that aligns attack semantics between provenance graphs and CTI reports at both coarse- and fine-grained levels through multi-objective pretraining, 3) a \textbf{CTI denoising module} that enhances the usability of multi-sourced CTI reports by utilizing LLMs' step-by-step reasoning capabilities; and (4) a \textbf{threat hunting module} that employs a two-stage retrieval mechanism to balance efficiency and precision in threat detection.
This section outlines the \textsc{APT-CGLP} architecture, as illustrated in Figure~\ref{framework}. We first introduce the Graph2CTI module that generates high-quality provenance graph-CTI report pairs for cross-modal supervision ($\S$\ref{sec:data_gen}). Next, we describe the multi-objective training framework, designed to align cross-modal attack semantics across both global and fine-grained levels through complementary learning tasks ($\S$\ref{sec:train}). Following this, the CTI denoising module is discussed, aimed at refining multi-sourced reports to enhance operational utility ($\S$\ref{sec:cti-denoising}). Finally, we detail the threat hunting module, which incorporates a two-stage retrieval strategy to balance detection precision and computational efficiency at scale ($\S$\ref{sec:threat hunting module}).

\subsection{Graph2CTI Module} \label{sec:data_gen}
In contrast to the abundance of image-caption datasets, 
the threat hunting domain faces a notable scarcity of paired provenance graph-CTI report samples for training (Challenge C2).
We tackle this gap through two key insights. First, audit logs provide a holistic and fine-grained account of system activities, and have long served as a trusted basis for forensic analysis \cite{aptshield,nodoze,depimpact}, making them a reliable data source for reverse-generating CTI reports.
Second, unlike intrusion detection that aims to explicitly identify attack patterns, threat hunting emphasizes semantic alignment between behavior patterns described in CTI reports and those embedded in provenance graphs \cite{megrapt,provgsearcher}. This fundamental shift allows us to exploit the ubiquitous attack-free audit logs to synthesize high-quality training pairs for cross-modal supervision.
These observations motivate the design of our Graph2CTI module, which first samples semantically rich subgraphs from benign provenance graphs, and then converts them into structured interaction triplets that can be processed by LLMs to generate corresponding CTI reports.

\noindent\textbf{Subgraph Sampling.} \label{pg_gen}
For constructing paired training samples, we begin by sampling subgraphs representing different activity instances as references.
While random sampling is a straightforward approach, it often yields skewed subgraphs dominated by monotonous interactions (e.g., repetitive file reads), which lack the behavior diversity inherent to real-world APT scenarios \cite{kairos, trec} and thus hinder the model's ability to learn generalizable features.
To overcome this, we design a heuristic-driven sampling algorithm that prioritizes attack-relevant patterns. It performs depth-first traversal, guided by the following three heuristics:
\begin{itemize}[leftmargin=*]
  \item \textbf{Depth Limitation}: The traversal depth is restricted to 2-3 layers, corresponding to behavior paths of 5-7 steps—typical of realistic APT campaigns \cite{megrapt}.
  \item \textbf{Size Constraint}: Each subgraph is limited to 10-20 nodes, a range derived from extensive statistical analysis of CTI reports.
  \item \textbf{Diverse Interactions}: Subgraphs are encouraged to include diverse entity types (e.g., processes, files, sockets) to emulate multi-stage attacks, where adversaries may initiate processes, establish remote connections, and exfiltrate files \cite{conan}.
\end{itemize}
These heuristics ensure that sampled subgraphs retain behavior semantics resembling APTs, enabling the generation of informative training pairs. We provide sampling pseudocode in Appendix \ref{sampling_algorithm}.

\begin{comment}
\begin{figure}[] 
  \centering
  \includegraphics[width=3in]{figs/cti generation.pdf}
  \caption{The workflow of the Graph2CTI module.}
  \label{cti_gen_fig}
\end{figure}
\end{comment}

\noindent\textbf{Synthetic CTI Generation.} \label{cti_gen}
LLMs have demonstrated language understanding and generation capabilities on par with domain experts \cite{llmsurvey, llm_news}. We leverage this capability to translate sampled subgraphs into coherent CTI reports in an automated manner.

First, each subgraph is transformed into a set of interaction triplets, such as $\langle word.exe,$
$ read, payroll.docx \rangle$ in Figure \ref{motivation}, which can be readily processed by LLMs.
Second, we employ in-context learning \cite{gpt3} to guide the LLM in compiling these triplets into CTI reports.
Concretely, we craft a structured prompt containing three parts: 1) a \textbf{task description} that articulates the CTI generation objective, ensuring that the LLM comprehends the intended task;
2) an \textbf{in-context example} that demonstrates a specific mapping from triplets to the corresponding CTI report, enabling the LLM to internalize the transformation patterns;
and 3) an \textbf{input section} that supplies the sequence of triplets to be translated.
The full prompt is subsequently submitted to LLMs, which outputs a synthetic CTI report that captures the high-level attack narrative of the original subgraph. Prompt design details are included in Appendix~\ref{cti_generation_prompt}.
% These synthesized provenance graph–CTI report pairs are used to augment the real dataset, yielding a rich and diverse training corpus.

\begin{comment}
\begin{figure*}[]
  \centering
  \includegraphics[width=7in]{figs/contrastive learning motivation.pdf}
  \caption{Comparisons of cross-modal contrastive learning scenarios. (a) Image retrieval: powered by pre-trained models and pertinent captions. (b) Molecular retrieval: benefiting from pertinent molecular descriptors. (c) Citation network classification: facilitated by graph homogeneity. (d) CTI report retrieval: complicated by heterogeneous graphs and verbose CTI contents.}
  \label{contrastive learning motivation} 
\end{figure*}
\end{comment}
\begin{figure}[]
  \centering
  \includegraphics[width=2.7in]{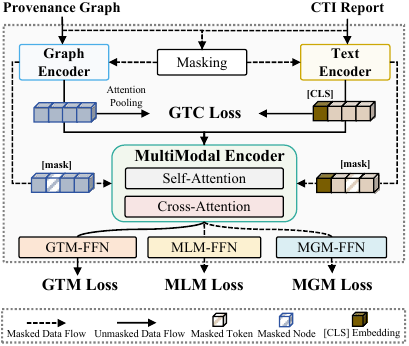}
  \caption{The training framework of \textsc{APT-CGLP}. GTC: graph-text contrastive, GTM: graph-text matching, MGM: masked graph modeling, MLM: masked language modeling, FFN: feed forward network.}
  %\caption{Training framework.}
  \label{training strategy}
\end{figure}

\subsection{Cross-Modal Training Module} \label{sec:train}
With ample paired training samples, an intuitive approach is to adopt cross-modal contrastive learning for semantic alignment, following the CLIP paradigm \cite{clip}. However, aligning provenance graphs with CTI reports presents domain-specific challenges.
First, unlike CLIP that benefits from powerful visual models pretrained on large-scale images, no such models exist for provenance graphs.
Second, CLIP focuses on coarse-grained global alignment \cite{sharegpt4v}, but APT campaigns often share similar high-level tactics (e.g., Collection, Exfiltration) \cite{conan}, making global alignment insufficient for fine-grained pattern differentiation.
Moreover, current graph-text contrastive methods \cite{g2p2, taga, graphtranslator} assume relatively homogeneous graphs and brief textual inputs, which are misaligned with our setting—where provenance graphs encode low-level, heterogeneous system events, and CTI reports contain high-level, unstructured behavior narratives (Challenge C1).
This semantic and structural asymmetry poses significant obstacles for cross-modal training.

To tackle these issues, we propose a training framework that bridges provenance graphs and CTI reports across both coarse- and fine-grained semantic levels.
As depicted in Figure \ref{training strategy}, the framework comprises three major components—a graph encoder, a text encoder, and a multimodal encoder—that are jointly optimized using a set of tailored objectives and task-specific heads to achieve cross-modal semantic alignment.
%First, contrastive learning aligns global embeddings of provenance graphs and CTI reports.
%Then, the multimodal encoder applies cross-attention on their node-level embeddings and token-level embeddings for fine-grained semantic matching.

\subsubsection{Encoders}
\textsc{APT-CGLP} integrates two unimodal encoders to independently process provenance graphs and CTI reports, along with a transformer-based multimodal encoder to learn their implicit correlations and produce unified representations.

\noindent\textbf{Graph Encoder.} 
To encode the structural interaction semantics of provenance graphs, we employ the Graph Isomorphism Network (GIN) \cite{gin}, denoted as $f_G$, for its powerful topological discrimination capacity.
Node attributes (see Table \ref{init_attr}) and edge types are initialized as graph features via our text encoder. Each provenance graph $G_i$ is then processed by $f_G$, which aggregates multi-hop neighborhoods for each node to encode its contextual behaviors (detailed in Appendix \ref{app:graph_encoder}). The resulting node embeddings are defined as:

\begin{equation}
\mathbf{H}_{G_i} = \{ \mathbf{h}_{n_1}, \mathbf{h}_{n_2}, \dots, \mathbf{h}_{n_{|V_i|}} \} =  f_G(G_i; \theta_G),
\end{equation}

\noindent 
where \( \mathbf{h}_{n_j} \in \mathbb{R}^d \) is the $j$-th node embedding, \( |V_i| \) is the number of nodes in \( G_i \), and \( \theta_G \) denotes the parameters of GIN.

\noindent\textbf{Text Encoder.}
We use BERT \cite{bert} as the text encoder $f_T$, due to its effectiveness in capturing bidirectional causal dependencies among system entities within CTI reports.
Each CTI report $T_i$ is first tokenized, and then passed through self-attention layers to produce token-level contextual embeddings:

\begin{equation}
  \mathbf{H}_{T_i} = \{ \mathbf{h}_{t_\text{[CLS]}}, \mathbf{h}_{t_1}, \mathbf{h}_{t_2}, \dots, \mathbf{h}_{t_{|T_i|}}\}  = f_T(T_i; \theta_T),
\end{equation}

\noindent
where $\mathbf{h}_{t_j} \in \mathbb{R}^d$ is the embedding  for token $t_j$, and $\mathbf{h}_{t_\text{[CLS]}}$ denotes the global semantic representation of the CTI report \cite{bert,albef}. The parameters of BERT are denoted by \(\theta_T\).

\begin{table}[]
  \scriptsize
  \centering
  \caption{Node attributes in provenance graphs.}
  \label{init_attr}
  \begin{tabular}{|l|l|l|}
  \hline
  \textbf{Node Type} & \textbf{Attribute} & \textbf{Example Instance} \\ \hline
  Process & Type: CommandLine & process: movingonup.exe fun.com 81 \\ \hline
  File & Type: Filepath & file: /142.20.61.135/share \\ \hline
  Socket & Type: IP Address:Port & socket: 127.0.0.1:0 \\ \hline
  \end{tabular}
\end{table}

\noindent\textbf{Multimodal Encoder.}
To learn semantic associations between provenance graphs and CTI reports, we introduce a multimodal encoder $f_M$, built on a transformer backbone. 
The encoder first applies self-attention over $\mathbf{H}_{T_i}$ to capture intra-text dependencies, and subsequently utilizes cross-attention mechanisms to integrate structured node semantics $\mathbf{H}_{G_i}$.
Such a design enables the model to learn detailed token-node interactions and achieve precise semantic alignment across modalities. The final multimodal representation is computed as:
\begin{equation}
  \mathbf{z}_m = f_M(\mathbf{H}_{G_i}, \mathbf{H}_{T_i} ; \theta_M),
  \end{equation}

\noindent where $\theta_M$ denotes the parameters of $f_M$, and $\mathbf{z}_m$ is the resulting joint representation for downstream alignment and threat hunting.

\subsubsection{Training Strategy}\label{sec:training strategy}
Global semantic alignment enforces coarse-grained consistency across modalities but lacks the sensitivity to differentiate attack campaigns with similar patterns. We mitigate this by adopting a hierarchical learning strategy: the global alignment objective establishes a fundamental cross-modal semantic coherence, while a set of auxiliary tasks offers fine-grained supervision to resolve subtle semantic inconsistencies.

\noindent \textbf{Graph-Text Contrastive Learning (GTC).}
This objective enforces global semantic alignment between provenance graphs and CTI reports by maximizing agreement between matched pairs while pushing apart mismatched ones in the shared embedding space.

Specifically, given a provenance graph-CTI report pair ($G_i,T_j$), the unimodal encoders $f_G$ and $f_T$ are used to extract their respective overall representations. The graph representation is derived via an attention pooling mechanism over node embeddings: $\mathbf{z}_{G_i} = \sum_{n \in V_i} \alpha_{n}\mathbf{h}_{n} $, where $\alpha_{n}$ is the attention weight of node $n$. The CTI representation $\mathbf{z}_{T_j}$ is extracted as the hidden state of its \texttt{[CLS]} token: $\mathbf{z}_{T_j} = \mathbf{h}_{t_\text{[CLS]}}$. 
We then compute the softmax-normalized similarity between the paired representations across a mini-batch as:
\begin{equation}
  p^{\text{g2t}}_{ij} (G, T) = \frac{\exp(sim(\mathbf{z}_{G_i}, \mathbf{z}_{T_j}) / \tau)}{\sum_{k=1}^{B} \exp(sim(\mathbf{z}_{G_i}, \mathbf{z}_{T_k}) / \tau)},
\end{equation}
where $sim(\cdot, \cdot)$ denotes dot-product similarity, $\tau$ is a learnable temperature parameter, and $B$ is the batch size.
Using one-hot supervision $\mathbf{y}^{\text{g2t}} \in \{0,1\}^{B\times B}$, where $\mathbf{y}^{\text{g2t}}_{ij} = 1$ indicates a positive pair, we optimize the model with a cross-entropy loss:
\begin{equation}
  \text{H}\left(\mathbf{y}^{\text{g2t}}, \mathbf{p}^{\text{g2t}}(G,T)\right) = - \sum_{i=1}^{B} \sum_{j=1}^{B} \mathbf{y}_{ij}^{\text{g2t}} \log \left( p_{ij}^{\text{g2t}}(G,T) \right),
\end{equation}
\begin{equation}
  \mathcal{L}_{\text{g2t}} = \mathbb{E}_{\{G, T\} \sim \mathcal{D}} \, \text{H}\left(\mathbf{y}^{\text{g2t}}, \mathbf{p}^{\text{g2t}}(G,T)\right),
\end{equation}
where $\{G, T\} \sim \mathcal{D}$ denotes a mini-batch sampled from the dataset $\mathcal{D}$, and $\mathcal{L}_{\text{g2t}}$ is the graph-to-text contrastive loss.
To enforce bidirectional semantic agreement, we compute a symmetric loss $\mathcal{L}_{t2g}$ for text-to-graph matching, and define the total GTC loss as: $\mathcal{L}_{gtc} = (\mathcal{L}_{g2t} + \mathcal{L}_{t2g}) / 2$.

\noindent \textbf{Graph-Text Matching (GTM).} To capture fine-grained semantic alignment, we introduce a binary matching task that predicts whether the behavior semantics of a given pair are consistent.
The multimodal encoder $f_M$ produces a joint embedding $\mathbf{z}_m$ that captures detailed interactions between graph nodes and CTI tokens. The $\mathbf{z}_m$ is then passed through a feedforward network (GTM-FFN) to compute the matching probability. The GTM loss is defined as: 
\begin{equation}
  \mathcal{L}_{\text{gtm}} = \mathbb{E}_{\{G, T\} \sim \mathcal{D}} \, \text{H}(\mathbf{y}^{\text{gtm}}, \mathbf{p}^{\text{gtm}}(G, T)),
\end{equation}
where $\textbf{y}^{\text{gtm}}$ denotes ground-truth labels. To enhance learning efficiency, a \textit{hard negative sampling} strategy \cite{albef} is employed to select non-matching pairs with high global similarities as challenging negatives, which sharpens the model's discriminative capacity.

\noindent \textbf{Masked Language Modeling (MLM).}
This objective strengthens cross-modal understanding by training the model to recover masked tokens in a corrupted CTI report $\hat{T}$ using both the visible text and the associated graph context.
To construct $\hat{T}$, we randomly replace a subset of tokens in the original CTI report with the special token \texttt{[MASK]}. Then, we feed ($\hat{T},G$) into $f_M$ to integrate structural and textual information via cross-attention, and a task-specific head (MLM-FFN) outputs the vocabulary distribution probability $\mathbf{p}^{\text{mlm}}(G, \hat{T})$ of masked tokens. Let $\mathbf{y}^{\text{mlm}}$ be the corresponding one-hot encoded ground-truth labels. The MLM loss is formulated as:
\begin{equation}
  \mathcal{L}_{\text{mlm}} = \mathbb{E}_{\{G, \hat{T}\} \sim \mathcal{D}} \, \text{H}(\mathbf{y}^{\text{mlm}}, \mathbf{p}^{\text{mlm}}(G, \hat{T})).
\end{equation}

\noindent \textbf{Masked Graph Modeling (MGM).}
Following the same principle as MLM, this task requires the model to reconstruct masked nodes in the provenance graph $\hat{G}$ based on the unmasked nodes and the paired CTI report.
Specifically, we randomly mask nodes in the original provenance graph to construct $\hat{G}$. Then, $f_M$ applies cross-attention between nodes in $\hat{G}$ and tokens in the paired CTI report $T$, followed by a prediction head (MGM-FFN) to predict masked node embeddings $\mathbf{p}^{\text{mgm}}(\hat{G},T)$. Let $\mathbf{y}^{\text{mgm}}$ represent the corresponding original node features. The training objective minimizes the mean squared error between the predicted and original node embeddings:
\begin{equation}
  \mathcal{L}_{\text{mgm}} = \mathbb{E}_{\{\hat{G}, T\} \sim \mathcal{D}} \left( \frac{1}{B} \sum_{i=1}^{B} ||\mathbf{y}_i^{\text{mgm}} - \mathbf{p}_i^{\text{mgm}}(\hat{G}, T)||_2^2 \right),
\end{equation}

The final loss function integrates above objectives through a weighted combination as follows:
\begin{equation}
  \mathcal{L} = \alpha \mathcal{L}_{\text{gtc}} + (1\!-\!\alpha)(\mathcal{L}_{\text{gtm}} + \mathcal{L}_{\text{mlm}} + \mathcal{L}_{\text{mgm}}),
\end{equation}
where $\alpha \in (0,1)$ controls the trade-off between coarse- and fine-grained semantic alignment. A larger weight is placed on GTC to establish global semantic consistency, while the auxiliary tasks foster more nuanced cross-modal understanding.

\subsection{CTI Denoising Module} \label{sec:cti-denoising}
Rather than being clean and structured like the synthetic CTI reports used for training, real-world CTI reports often contain substantial noise, such as website metadata \cite{attackg}, which obscures the core attack insights essential for threat hunting (Challenge C3).
To mitigate this, we leverage the advanced language summarization capabilities of LLMs \cite{llm_news} to isolate and restructure actionable intelligence from raw CTI inputs, effectively filtering out extraneous content \cite{attackg+}. 
Particularly, we adopt a Chain-of-Thought (CoT) reasoning framework that decomposes the denoising process into logically ordered subtasks, emulating the analytical workflow of threat analysts.  
%This emulates the analytical workflow of security analysts, who iteratively identify key entities and infer contextual relationships from unstructured text.
The CoT prompt comprises three reasoning stages:
\begin{enumerate}[leftmargin=*]
  \item \textbf{Entity Identification:}  Scan the CTI report to extract core system entities (e.g., malicious payloads) involved in the attack.
  \item \textbf{Interaction Extraction:} Infer both explicit and implicit interactions among entities based on contextual cues.
  \item \textbf{Knowledge Distillation:} Consolidate the extracted interactions into a concise and temporally ordered attack narrative.
\end{enumerate}

By guiding the LLM through this structured reasoning process, we obtain distilled intelligence that retains essential attack patterns while suppressing noise. Complete prompt templates are provided in Appendix~\ref{cti_clean_instructions}.

\subsection{Threat Hunting Module}\label{sec:threat hunting module}

Effective threat hunting is enabled by integrating the trained multimodal encoder and GTM-FFN to recursively estimate the matching probabilities between provenance graphs and CTI reports (see Section \ref{sec:training strategy}).
However, directly applying this matching process over large-scale provenance graphs and an expanding CTI corpus is computationally prohibitive.
To address this scalability bottleneck, we reformulate the threat hunting as a \textit{two-stage retrieval pipeline}. The first stage narrows the search space via a lightweight similarity-based retrieval, while the second stage conducts fine-grained semantic matching for high-precision threat identification. 
\begin{enumerate}[leftmargin=*]
  \item \textbf{Coarse-Grained Retrieval:}
  All CTI reports are embedded into dense representations $\{\mathbf{z}_t^1, \mathbf{z}_t^2, \dots, \mathbf{z}_t^n\}$ using the text encoder and indexed in a vector database.
  Given a target provenance graph, the graph encoder computes its global embedding $\mathbf{z}_g$.
  Top-$k$ CTI candidates are then retrieved based on cosine similarity between $\mathbf{z}_g$ and stored CTI embeddings.
  \item \textbf{Fine-Grained Matching:}
  The retrieved CTI candidates and $\mathbf{z}_g$ are passed into the multimodal encoder to produce joint representations $\{\mathbf{z}_m^1, \mathbf{z}_m^2, \dots, \mathbf{z}_m^k\}$.
  These embeddings are then fed into the GTM-FFN, which computes matching probabilities, enabling precise identification of the most relevant CTI report.
  %If multiple reports are identified as matches, the report with the highest similarity score is selected.
\end{enumerate}
%-------------------------------------------------------------------------------
\section{Evaluation}
%-------------------------------------------------------------------------------
We conduct comprehensive experiments to investigate the following research questions (RQs):
\begin{itemize}[leftmargin=*]
  \item \textbf{RQ1:} How does \textsc{APT-CGLP} compare to state-of-the-art threat hunting systems?
  \item \textbf{RQ2:} What are the individual contributions of its core modules, namely training data augmentation, cross-modal training, and CTI denoising to overall performance?
  \item \textbf{RQ3:} How well does the two-stage retrieval strategy balance threat hunting precision and computational efficiency?
  \item \textbf{RQ4:} How effective is \textsc{APT-CGLP} in supporting alert validation?
\end{itemize}

\subsection{Experiment Settings}
This section describes the datasets, evaluation protocol, metrics, and the implementation and hardware configuration used in our experiments.

\subsubsection{Datasets}
We evaluate \textsc{APT-CGLP} on the DARPA TC E3 \cite{darpatce3} and OpTC \cite{darpatcoptc} datasets, both of which simulate real-world enterprise-scale APT campaigns.
The E3 dataset spans two weeks of system audit logs, with the first week recording benign activity and the second week containing various attack scenarios (e.g., Nginx exploits, Firefox backdoors).
We select the Cadets, Trace, and Theia subsets for evaluations due to their comprehensive audit coverage and well-documented reports.
The OpTC dataset is the latest iteration of the DARPA TC project, offering over 17 billion events across 1,000 Windows hosts. It similarly comprises multiple days of benign activities followed by three days of sophisticated attack scenarios, such as ``PowerShell Empire'' and ``Malicious Upgrade''. Audit logs generated from the victim hosts are utilized for evaluation.
%accompanied with detailed after-action reports for evaluation.
%We argue that leveraging these datasets facilitates a comprehensive assessment of \textsc{APT-CGLP}, given their diverse attack scenarios, extensive audit records, and reliable ground truth for validation.

\subsubsection{Evaluation Protocol} We partition datasets into training and testing subsets, then establish comprehensive evaluation metrics.

\noindent \textbf{Pre-training Strategy.}
We use only benign audit logs from the first week of each dataset for training. 
% of both the E3 and OpTC datasets. Concretely, 
The Graph2CTI module transforms daily logs into provenance graphs and samples representative subgraphs, which are then processed by the LLM to generate synthetic CTI reports (Section~\ref{cti_gen}). 
As summarized in Table \ref{dataset distributions}, this process yields 45,225 provenance graph-CTI report pairs.
Crucially, attack-related data is excluded during training data curation to prevent \textit{data leakage} during evaluation \cite{dosdont}.

\noindent \textbf{Testing Strategy.}
For rigorous assessment, we curate a hybrid test corpus containing both malicious and benign provenance graphs, along with a large collection of web-sourced CTI reports. 
\begin{itemize}[leftmargin=*]
  \item \textbf{Provenance Graphs.} 
  We evaluate two types of subgraphs: \textit{malicious subgraphs} for attack detection and \textit{benign subgraphs} for assessing false alert filtering.
  All subgraphs are generated using the MEGR-APT sampling strategy \cite{megrapt}. First, anomalous entities are selected as seed nodes based on predefined heuristics.
  For malicious cases, seeds are entities with names matching IoCs and interaction timestamps aligning with attack windows recorded in DARPA reports. This cross-validation guarantees accurate annotation.
  For benign cases, seeds are derived from: 1) entities in \textit{benign logs} matching IoC names, and 2) false alarms reported by the APT detection system Magic \cite{magic}. Anchored on each seed node, subgraphs are then constructed by expanding to include neighboring seeds and process nodes within a 2-3-hop radius.

  \item \textbf{CTI Corpus.}  The CTI corpus composes two sources: official DARPA reports and web-sourced reports from security vendors (e.g., MITRE \cite{attck}, VirusTotal \cite{virustotal}, Microsoft \cite{microsoft}). 
  To ensure quality and relevance, we apply regex-based filtering to an initial pool of 10,618 reports, retaining 5,172 CTI reports that contain detailed IoC and TTP descriptions. Keyword analysis confirms that these reports cover attack vectors such as phishing, living-off-the-land, and backdoor exploits. All CTI reports are processed through our denoising module to distill core insights, forming the external knowledge base for threat hunting.
\end{itemize}
Table~\ref{dataset distributions} presents the final testing dataset, which consists of 2,402 subgraphs and 5,183 CTI reports. Among them, 2,094 subgraphs are generated by Magic and are used for alert validation evaluation, while the remaining subgraphs (with 1.2\% malicious) are generated by MEGR-APT and used for threat hunting evaluation.
\begin{table}[]
  \scriptsize
  \centering
  \caption{Training and testing datasets. PG: provenance graph.}
  \label{dataset distributions}
  \begin{tabular}{|c|cc|ccc|}
    \hline
    \multirow{3}{*}{\textbf{Dataset}} & \multicolumn{2}{c|}{\textbf{Pre-training}} & \multicolumn{3}{c|}{\textbf{Testing}} \\ \cline{2-6} 
    & \multicolumn{1}{c|}{\multirow{2}{*}{\textbf{\# PG}}} & \multirow{2}{*}{\textbf{\# CTIs}} & \multicolumn{1}{c|}{\multirow{2}{*}{\textbf{\# PG}}} & \multicolumn{2}{c|}{\textbf{\# CTIs}} \\ \cline{5-6} 
    & \multicolumn{1}{c|}{} &  & \multicolumn{1}{c|}{} & \multicolumn{1}{c|}{\textbf{\# DARPA}} & \textbf{\# Web-Sourced} \\ \hline
   E3-CADETS & \multicolumn{2}{c|}{7,524} & \multicolumn{1}{c|}{721} & \multicolumn{1}{c|}{4} & \multirow{4}{*}{5,172} \\
   E3-THEIA & \multicolumn{2}{c|}{10,263} & \multicolumn{1}{c|}{318} & \multicolumn{1}{c|}{2} &  \\
   E3-Trace & \multicolumn{2}{c|}{20,781} & \multicolumn{1}{c|}{873} & \multicolumn{1}{c|}{2} &  \\
   OpTC & \multicolumn{2}{c|}{6,657} & \multicolumn{1}{c|}{490} & \multicolumn{1}{c|}{3} &  \\ \hline
   \textbf{Total} & \multicolumn{2}{c|}{45,225} & \multicolumn{1}{c|}{2,402} & \multicolumn{2}{c|}{5,183} \\ \hline
  \end{tabular}
  \end{table}

\subsubsection{Evaluation Metrics.}
We define a positive match between a provenance graph and a CTI report when: 1) their unimodal embedding similarity exceeds threshold $\lambda$, and 2) GTM-FFN predicts a match based on multimodal embeddings. Otherwise, it is a negative match.
True positives occur when malicious subgraphs match their paired CTI report \textit{exclusively}, whereas false negatives indicate unmatched malicious subgraphs.
True negatives represent benign subgraphs with no CTI matches, while false positives denote benign subgraphs incorrectly matched to CTI reports.
Our evaluations use these definitions with standard metrics: Recall, Precision, Accuracy, False Positive Rate (FPR), and F1-Score.

\subsubsection{Environment and Implementation} \label{app:imp}
All experiments were conducted on an Ubuntu 20.04 server equipped with an Intel Xeon Gold 5128 CPU, 128 GB of RAM, and two NVIDIA RTX 4090 GPUs.

The implementation of \textsc{APT-CLIP} comprises $\sim$5,000 lines of Python codes.
We adopt the BERT-Base model \cite{bert-base} as a pre-trained text encoder and fine-tune the last two layers. The graph encoder is implemented as a three-layer GIN network with its dimensionality aligned to that of the text encoder. A custom transformer-like model is designed as the multimodal encoder for cross-modal integration.
During training, the token masking ratio for the MLM objective is set to 15\%, and one node was randomly masked for the MGM objective.
Optimization is performed using AdamW with a learning rate of $2\times10^{-4}$ and a weight decay of 0.01. 
A cosine annealing schedule is applied over 100 epochs, decaying the learning rate to a minimum of $1\times10^{-5}$.
During the first 7 epochs, a warmup phase is employed.
The loss-balancing weight is set to $\alpha=0.7$.
To mitigate overfitting, regularization and dropout strategies were implemented during model training.
The vector database for storing and retrieving CTI report embeddings is based on FAISS \cite{faiss}. 
The optimal number of candidate CTI in the two-stage retrieval strategy was fixed at $k=10$, and the similarity threshold was set to $\lambda=0.5$.
Finally, the open-source LLaMA3-8B model~\cite{llama3} is deployed locally for synthetic training data generation and CTI report denoising, providing lower computational overhead and improved security when processing sensitive intelligence data.

\begin{table*}[!]
  \scriptsize
\setlength{\tabcolsep}{3.5pt}
\centering
  \caption{Comparisons of \textsc{APT-CGLP} with state-of-the-art threat hunting methods across DARPA datasets. Best results are bolded, and second-best are underlined. $\uparrow$: higher is better. $\downarrow$: lower is better. Rec.:Recall, Pre.: Precision, Acc.:Accuracy, F1.:F1-Score.}
  \label{th_comp}
  \begin{tabular}{l|ccccc|ccccc|ccccc|ccccc}
    \hline
    \textbf{Dataset} & \multicolumn{5}{c|}{\textbf{E3-Cadets}} & \multicolumn{5}{c|}{\textbf{E3-Theia}} & \multicolumn{5}{c|}{\textbf{E3-Trace}} & \multicolumn{5}{c}{\textbf{OpTC}} \\ \hline
    \textbf{Model} & \textbf{Rec. $\uparrow$} & \textbf{FPR $\downarrow$} & \textbf{Pre.$\uparrow$} & \textbf{Acc.$\uparrow$} & \textbf{F1.$\uparrow$} & \textbf{Rec. $\uparrow$} & \textbf{FPR $\downarrow$} & \textbf{Pre.$\uparrow$} & \textbf{Acc.$\uparrow$} & \textbf{F1.$\uparrow$}  & \textbf{Rec. $\uparrow$} & \textbf{FPR $\downarrow$} & \textbf{Pre.$\uparrow$} & \textbf{Acc.$\uparrow$} & \textbf{F1.$\uparrow$} & \textbf{Rec. $\uparrow$} & \textbf{FPR $\downarrow$} & \textbf{Pre.$\uparrow$} & \textbf{Acc.$\uparrow$} & \textbf{F1.$\uparrow$}  \\ \hline
    \begin{tabular}[c]{@{}c@{}}ProvG-Searcher w/o HI\end{tabular} & 0.538 & 0.212 & 0.389 & 0.738 & 0.452 & 0.643 & 0.226 & 0.429 & 0.746 & 0.514 & 0.429 & 0.219 & 0.115 & 0.759 & 0.182 & 0.381 & 0.264 & 0.222 & 0.677 & 0.281 \\
    \begin{tabular}[c]{@{}c@{}}MEGR-APT w/o HI\end{tabular} & 0.615 & 0.192 & 0.444 & 0.769 & 0.516 & 0.786 & 0.132 & 0.611 & 0.851 & 0.688 & 0.571 & 0.152 & 0.200 & 0.830 & 0.296 & 0.571 & 0.179 & 0.387 & 0.780 & 0.462 \\
%==================================
\begin{tabular}[c]{@{}c@{}}TREC w/o HI\end{tabular} & 0.692 & 0.173 & 0.500 & 0.799 & 0.581 & 0.714 & 0.151 & 0.556 & 0.821 & 0.625 & 0.714 & 0.157 & 0.233 & 0.835 & 0.351 & 0.524 & 0.170 & 0.380 & 0.780 & 0.441 \\\hline
%==================================
    \begin{tabular}[c]{@{}c@{}}ProvG-Searcher w/ HI\end{tabular} & 1.000 & 0.135 & 0.650 & 0.892 & 0.788 & 1.000 & 0.113 & 0.700 & 0.910 & 0.824 & 1.000 & 0.056 & 0.636 & 0.949 & 0.778 & {\underline{1.000}} & 0.085 & 0.700 & 0.929 & 0.824 \\
    \begin{tabular}[c]{@{}c@{}}MEGR-APT w/ HI\end{tabular} & {\underline{1.000}} & {\underline{0.058}} & {\underline{0.813}} & {\underline{0.954}} & {\underline{0.897}} & 1.000 & \textbf{0.038} & \textbf{0.875} & \textbf{0.970} & \textbf{0.933} & 1.000 & 0.028 & 0.778 & 0.974 & 0.875 & \textbf{1.000} & {\underline{0.057}} & {\underline{0.778}} & {\underline{0.953}} & {\underline{0.875}}\\
%==================================
\begin{tabular}[c]{@{}c@{}}TREC w/ HI\end{tabular} & 1.000 & 0.077 & 0.765 & 0.938 & 0.877 & {\underline{1.000}} & 0.075 & 0.778 & 0.941 & 0.875 & {\underline{1.000}} & {\underline{0.021}} & {\underline{0.824}} & {\underline{0.981}} & {\underline{0.903}} & 1.000 & 0.075 & 0.724 & 0.937 & 0.840 \\ \hline
%==================================
    \textsc{APT-CGLP} & \textbf{1.000} & \textbf{0.019} & \textbf{0.929} & \textbf{0.985} & \textbf{0.963} & \textbf{1.000} & {\underline{0.057}} & {\underline{0.824}} & {\underline{0.955}} & {\underline{0.903}} & \textbf{1.000} & \textbf{0.014} & \textbf{0.875} & \textbf{0.987} & \textbf{0.933} & 0.952 & \textbf{0.038} & \textbf{0.833} & \textbf{0.961} & \textbf{0.889} \\ \hline
    \end{tabular}
  \end{table*}

\subsection{RQ1: Threat Hunting Performance}
We compare \textsc{APT-CGLP} against three state-of-the-art baselines:
\begin{itemize}[leftmargin=*]
  \item \textbf{MEGR-APT} \cite{megrapt}: Learns attack graph representations for both query and provenance graphs using abstract node and edge features, followed by a Neural Tensor Network \cite{ntn} for alignment.
  \item \textbf{ProvG-Searcher} \cite{provgsearcher}: Formulates threat hunting as a subgraph entailment problem using GNNs and order embeddings \cite{basepaper}, with heuristic node abstractions for feature augmentation.
  \item \textbf{TREC} \cite{trec}: A heterogeneous graph attention network \cite{han} encodes the behavior features of provenance subgraphs and CTI-derived query graphs into a shared embedding space via a siamese architecture, enabling few-shot recognition of similar patterns.
\end{itemize}

All baselines operate on CTI-derived query graphs for threat hunting. To ensure a rigorous comparison, we construct two variants for each: 1) the \textbf{w/o HI} variant utilizes query graphs automatically extracted from denoised CTI reports using AttacKG \cite{attackg}, simulating a fully automated pipeline without human involvement (HI); 2) the \textbf{w/ HI} variant uses manually refined query graphs with complete semantics, consistent with their original implementations and reflective of state-of-the-art performance.
Evaluation results are presented in Table~\ref{th_comp}, with key findings summarized as follows:
\begin{itemize}[leftmargin=*]
  \item \textbf{Impact of Human Involvement.}
  The w/o HI variants suffer from clear performance drops. On the E3-Cadets dataset, TREC w/o HI achieves 0.581 F1-Score, whereas incorporating human-in-the-loop intervention boosts the score to 0.877—a notable improvement of 29.6\%. Similarly, ProvG-Searcher jumps from 0.452 (w/o HI) to 0.788 (w/ HI). This trend is consistent across datasets, confirming that semantic gaps in auto-generated query graphs degrade detection capability. The performance disparity highlights current CTI parsing limitations and the continued need for human curation to preserve semantic fidelity.
  \item \textbf{Superiority of \textsc{APT-CGLP}.}
  \textsc{APT-CGLP} achieves consistently strong performance across all datasets, setting a new benchmark for fully automated systems. Notably, it surpasses the w/o HI baselines by over 30\% in F1-Score. The Recall for \textsc{APT-CGLP} is also perfect across three datasets and maintains lower FPRs. These results demonstrate its ability to capture both holistic consistent attack semantics automatically and robustly, outperforming baselines even with handcrafted query graphs.
  \item \textbf{Comparison with Human-Tuned Baselines.} Remarkably, \textsc{APT-CGLP} matches or exceeds the performance of w/ HI baselines on most datasets. For instance, it improves F1-Score upon MEGR-APT (w/ HI) by 6.6\% on E3-Cadets and by 5.8\% on E3-Trace. The only dataset where it slightly lags is E3-Theia, likely due to ambiguous CTI descriptions that cause several matches with normal provenance graphs. However, its FPR remains low (0.057), suggesting these errors are well-contained. Overall, \textsc{APT-CGLP} offers a high-performance, scalable alternative to manually curated systems, significantly lowering operational cost.
\end{itemize}
%that effectively captures both global and fine-grained semantic similarities between provenance graphs and CTIs, thereby enhancing overall threat hunting performance and efficiency.

\subsection{RQ2: Ablation Studies} \label{sec:ablation}
\begin{figure}[]
  \centering
  \includegraphics[width=3.3in]{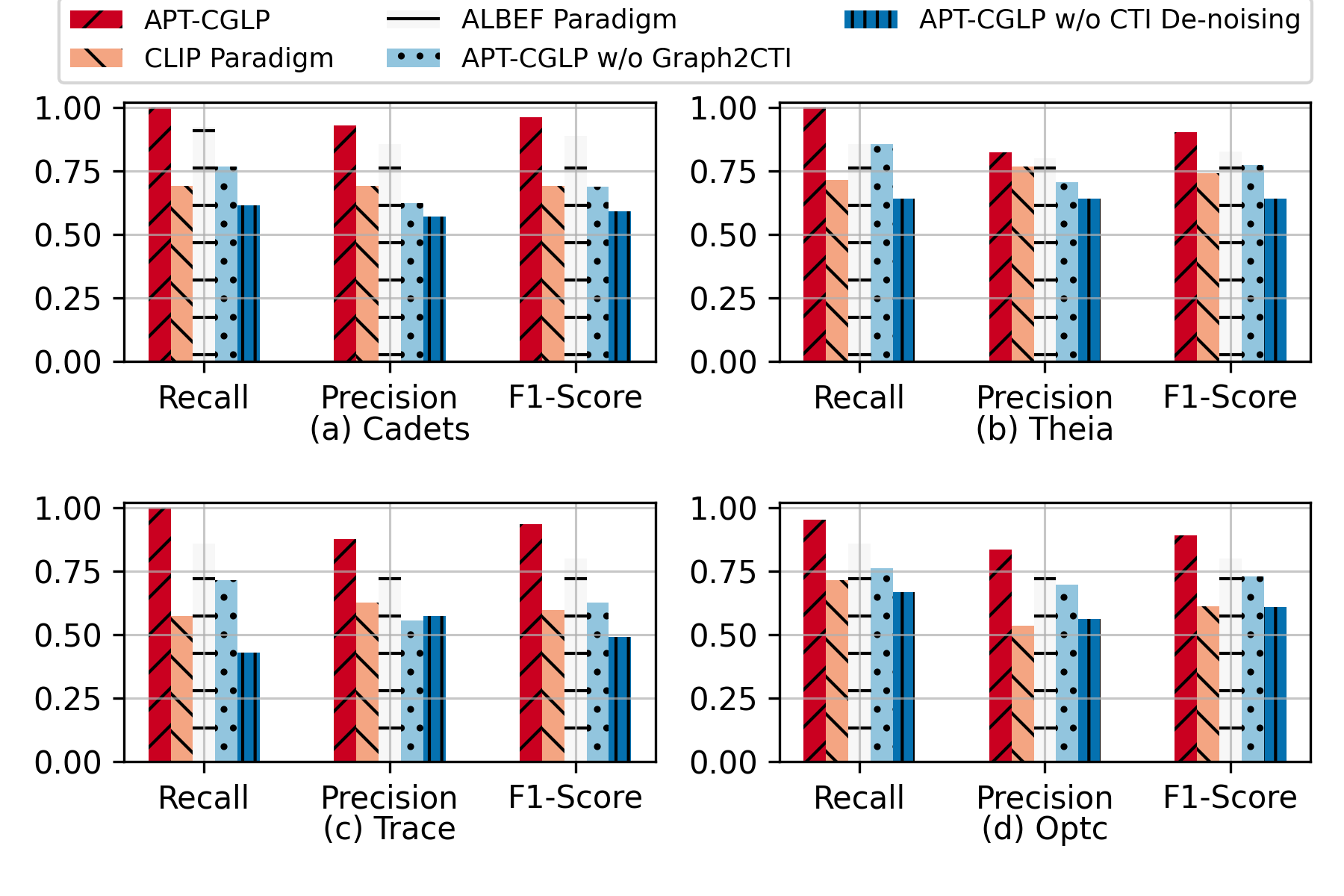}
  \caption{Results of ablation studies across different datasets.}
  \label{ablation}
\end{figure}

We conduct ablation studies to evaluate the contributions of each component—training data augmentation, different cross-modal training strategies, and CTI denoising—on the overall threat hunting performance of \textsc{APT-CGLP}. Results are presented in Figure~\ref{ablation}.

\noindent\textbf{Effect of Training Data Augmentation.}
We assess the efficacy of Graph2CTI module by comparing it to a variant that uses naive random subgraph sampling (\textsc{APT-CGLP} w/o Graph2CTI) during training.
As illustrated in Figure \ref{ablation}, removing Graph2CTI leads to over 10\% performance drop, likely due to the lack of semantic diversity and behavior richness in randomly sampled graphs. 
%These results highlight the critical role of optimized training data in enabling the model to effectively capture multi-dimensional semantic correlations between provenance graph structures and attack descriptions during training, thereby improving its generalization to real-world scenarios.
This confirms that training with informative subgraphs—mirroring real-world attack patterns—is critical for model generalization.

\noindent\textbf{Efficacy of Fine-Grained Semantic Alignment.}
We evaluate the effect of fine-grained alignment by comparing \textsc{APT-CGLP} with a CLIP-style variant that only performs global contrastive learning at the embedding space (the CLIP paradigm).
This variant yields the second-lowest F1-Score across datasets, highlighting that global alignment alone is insufficient to bridge the structural and semantic modality gap. In contrast, our model's use of a multimodal encoder with cross-attention achieves over 20\% gains in F1-Score, validating the necessity of modeling fine-grained node-token interactions. %global alignment alone is insufficient given the modality gap and granular nature of CTI data; 

%These results suggest that the global semantic alignment is less effective for threat hunting due to the substantial modality gap between provenance graphs and CTI reports, as well as the inherent complexity of APT attacks and detail-oriented nature of CTI reports. Therefore, capturing fine-grained feature interactions through cross-attention can uncover detailed cross-modal correlations, thus improving the threat hunting performance.

\noindent\textbf{Role of Inter-Modal Masked Modeling.}
To evaluate the effectiveness of inter-modal masked modeling strategy, we compare \textsc{APT-CGLP} with ALBEF \cite{albef}, a prior framework that uses only masked language modeling for vision-language alignment (the ALBEF paradigm). 
Figure \ref{ablation} indicates that \textsc{APT-CGLP} consistently outperforms ALBEF in F1-Scores. 
These findings underscore the essential role of bidirectional cross-modal supervision for effectively capturing subtle semantic cues critical to robust threat hunting.
%, which is critical for robust threat identification.

\noindent\textbf{Influence of CTI Denoising.}
We assess the utility of the CTI denoising module by comparing \textsc{APT-CGLP} to a variant using noisy CTI reports (\textsc{APT-CGLP} w/o CTI denoising).
Figure \ref{ablation} shows that denoising consistently improves Recall by over 30\% across all datasets.
This substantial improvement confirms that noisy content significantly hinders CTI utilization. By distilling concise, attack-relevant descriptions, both precision and robustness are improved.
%the denoising module enhances the model's ability to match CTI semantics with provenance structures, improving both precision and robustness.
% reinforcing its suitability for deployment in dynamic threat environments.
%These findings highlight the critical importance of precise and high-quality CTI reports in generating meaningful multimodal representations, ultimately improving downstream retrieval tasks in threat hunting.

\subsection{RQ3: Practicality of Retrieval Strategy}

\noindent\textbf{Effectiveness Analysis.}
We assess the effectiveness of the proposed two-stage retrieval strategy by comparing it with a similarity-based baseline and examining the influence of the candidate pool size $k$.
The baseline approach retrieves the CTI report with the highest similarity score as the threat hunting result. %\footnote{The similarity threshold is selected to maximize F1-Score.} 
For \textsc{APT-CGLP}, we assess two configurations with $k = 10$ and $k = 20$.
The experimental results in Table~\ref{retrieval comp} yield the following observations:
\begin{itemize}[leftmargin=*]
  \item \textbf{Improved Precision via Two-Stage Retrieval.} \textsc{APT-CGLP} consistently outperforms the similarity-based baseline across all benchmarks. The performance gains stem from the second-stage refinement, which enables fine-grained semantic alignment and effectively reduces false matches with benign subgraphs. This hierarchical design boosts Recall and overall robustness.
  \item \textbf{Candidate Pool Size Trade-offs.} Increasing the candidate size $k$ from 10 to 20 improves Recall (e.g., on the OpTC dataset) by broadening the search space. However, this also leads to a marginal increase in FPRs due to more semantically relevant candidates that are mistakenly matched. These findings suggest that $k$ should be adjusted according to deployment requirements.
  %: higher values favor high-Recall applications, while smaller values are preferable when minimizing overhead or false positives is critical.
\end{itemize}

\begin{table}[]
  \scriptsize
  \centering
  \caption{Comparison of similarity-based retrieval and two-stage retrieval strategies in threat hunting. Retr.: retrieval.}
  \label{retrieval comp}
  \begin{tabular}{c|c|c c c c c}
    \hline
    \textbf{Dataset} & \textbf{Strategy} &\textbf{Rec. $\uparrow$} & \textbf{FPR $\downarrow$} & \textbf{Pre.$\uparrow$} & \textbf{Acc.$\uparrow$} & \textbf{F1.$\uparrow$} 
    \\ \hline 
    \multirow{3}{*}{E3-Cadets} & Similarity-based Retr. & 0.923 & 0.058 & 0.800 & 0.938 & 0.857 \\
    & Two-stage Retr. ($k$@10) & \textbf{1.000} & \textbf{0.019} & \textbf{0.929} & \textbf{0.985} & \textbf{0.963} \\
    & Two-stage Retr. ($k$@20) & {\underline{1.000}} & {\underline{0.038}} & {\underline{0.867}} & {\underline{0.969}} & {\underline{0.929}} 
    \\ \hline 
   \multirow{3}{*}{E3-Theia} & Similarity-based Retr. & 0.929 & {\underline{0.075}} & {\underline{0.765}} & 0.925 & 0.839 \\
    & Two-stage Retr. ($k$@10) & \textbf{1.000} & \textbf{0.057} & \textbf{0.824} & \textbf{0.955} & \textbf{0.904} \\
    & Two-stage Retr. ($k$@20) & {\underline{1.000}} & 0.094 & 0.737 & {\underline{0.925}} & {\underline{0.849}}     
    \\ \hline 
   \multirow{3}{*}{E3-Trace} & Similarity-based Retr. & 1.000 & 0.042 & 0.700 & 0.962 & 0.824 \\
    & Two-stage Retr. ($k$@10) & \textbf{1.000} & \textbf{0.014} & \textbf{0.875} & \textbf{0.987} & \textbf{0.933} \\
    & Two-stage Retr. ($k$@20) & {\underline{1.000}} & {\underline{0.028}} & {\underline{0.778}} & {\underline{0.974}} & {\underline{0.875}} 
    \\ \hline 
   \multirow{3}{*}{OpTC} & Similarity-based Retr. & 0.905 & 0.094 & 0.655 & 0.906 & 0.760 \\
    & Two-stage Retr. ($k$@10) & {\underline{0.952}} & \textbf{0.038} & \textbf{0.833} & \textbf{0.961} & \textbf{0.889} \\
    & Two-stage Retr. ($k$@20) & \textbf{1.000} & {\underline{0.066}} & {\underline{0.750}} & {\underline{0.945}} & {\underline{0.857}} \\ \hline
   \end{tabular}
    \end{table}
\noindent\textbf{System Overhead.} \label{sec:overhead}
We further assess the computational overhead of \textsc{APT-CGLP} using the E3-Cadets dataset, focusing on memory usage and inference latency under varying candidate sizes $k\in\{10,20,30,40,50\}$ across both CPU and GPU environments.

\begin{figure}[]
  \centering
  \subfloat[\small{Inference memory cost.}]{\includegraphics[width=1.5in]{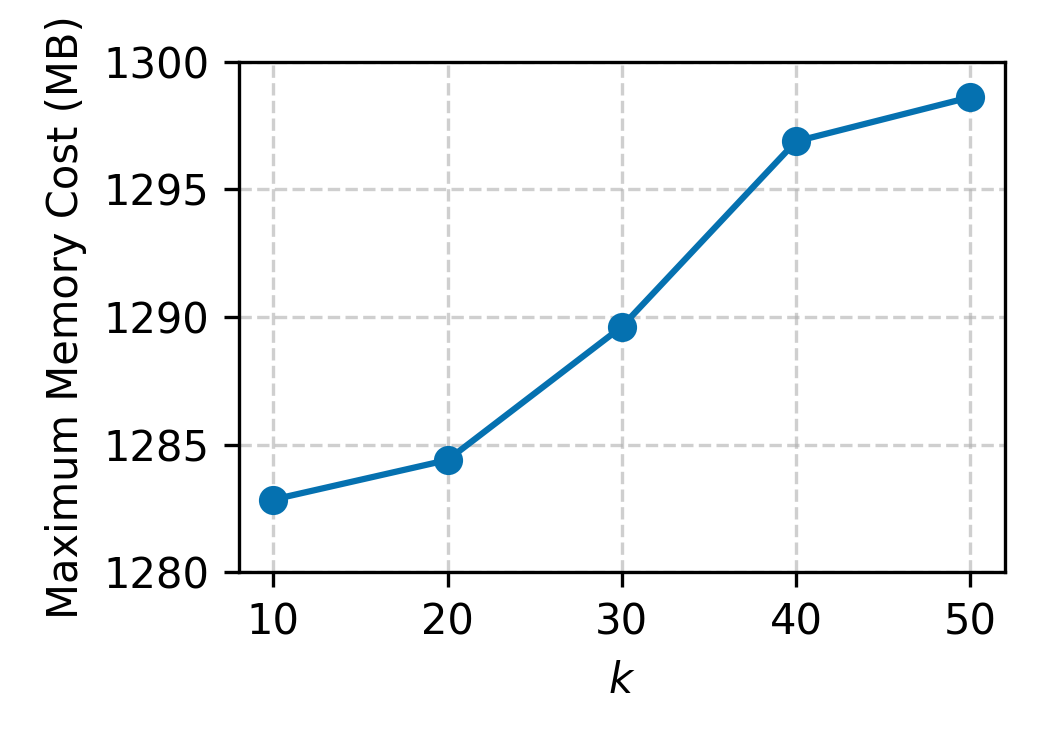}%
  \label{overhead_size}}
  \hfil
  \subfloat[\small{Inference time cost.}]{\includegraphics[width=1.5in]{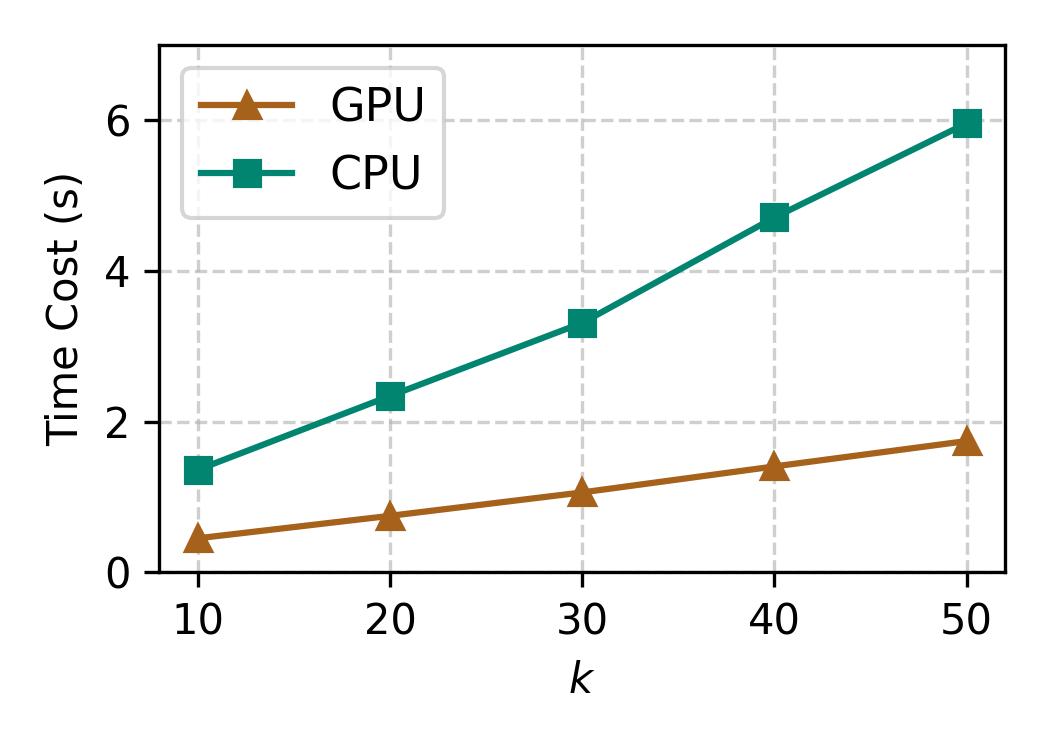}%
  \label{overhead_time}}
  \caption{Overhead of \textsc{APT-CGLP} for different $k$.}
  \label{fig_sim}
  \end{figure}

\begin{itemize}[leftmargin=*]
  \item \textbf{Memory Cost.} The total memory cost consists of (1) offline storage for CTI embeddings within a vector database (e.g., FAISS), and (2) runtime memory for the two-stage retrieval pipeline. Offline storage scales linearly with the corpus size, requiring approximately 0.5 MB per 1,000 CTI embeddings. As shown in Figure~\ref{overhead_size}, runtime memory is dominated by model weights and external dependencies, remaining stable at approximately $\sim$1 GB across all settings. Increasing $k$ raises a marginal memory cost. For example, scaling $k$ from 10 to 50 increases memory by just 16 MB, indicating strong memory efficiency and scalability.
  \item \textbf{Time Cost.}  The time cost includes (1) vector-based retrieval of top-$k$ CTI candidates, and (2) fine-grained multimodal matching for threat hunting.
  As illustrated in Figure~\ref{overhead_time}, latency grows linearly with $k$, but remains manageable. At $k=50$, a single threat query completes in $\sim$1.7 seconds on GPU and $\sim$6 seconds on CPU. Reducing $k$ to 10 lowers the GPU time to 0.4 seconds with negligible performance degradation.
  Overall, \textsc{APT-CGLP} supports real-time operation at scale. For instance, an enterprise-scale alert stream ($\sim$2,000 events \cite{nodoze}) can be processed in under 15 minutes, underscoring the framework's practicality.
\end{itemize}

\subsection{RQ4. Alert Validation Analysis}
\textsc{APT-CGLP} can also serve as an alert validator to reduce false positives generated by upstream APT detection systems. To evaluate this capability, we leverage alerts produced by the anomaly-based detection system MAGIC \cite{magic}, including both true and false alarms.
Moreover, two metrics are introduced to quantify alert verification performance: 1) \textbf{Alert Filtering Rate (AFR)}—the proportion of total alerts that are successfully filtered; and 2) \textbf{Threat Retention Rate (TRR)}—the fraction of true threat alerts correctly preserved.

As reported in Table~\ref{fp_filter}, \textsc{APT-CGLP} achieves AFRs over 90\% across all four datasets, while maintaining consistently high TRRs. These results demonstrate its strong capability to eliminate spurious alerts without compromising coverage of genuine threats.
Anomaly-based detectors \cite{threatrace, kairos, aptkgl, magic} often exhibit elevated FPRs due to rigid thresholds and reliance on behavior baselines, exacerbating the problem of alert fatigue \cite{nodoze} that delays or impairs incident response.
As benign behavior evolves over time, these systems often misclassify normal deviations as threats \cite{depimpact}. 
We mitigate this issue by grounding alert filtering using external CTI knowledge. 
%Through automated comparisons between system alerts and known attack patterns, it significantly reduces the manual burden on analysts and provides a practical solution for real-world threat environments.
  
\begin{table}[!t]
  \scriptsize
  \centering
  \caption{The performance gains generated by \textsc{APT-CGLP} through alert validation.}
  \label{fp_filter}
  \begin{tabular}{c|cccc}
  \hline
  \textbf{Gains} & \textbf{E3-Cadets} & \textbf{E3-Theia} & \textbf{E3-Trace} & \textbf{OpTC} \\ \hline
  AFR $\uparrow$ & 0.981 & 0.943 & 0.986 & 0.962 \\ \hline
  TRR $\uparrow$ & 1.000 & 1.000 & 1.000 & 1.000 \\ \hline
  \end{tabular}
\end{table}
\section{Discussion}
\noindent\textbf{LLM Generalization}: We tested various LLMs (e.g., Qwen2.5) and found that the proposed framework is not tied to a specific LLM. This is because the tasks performed in our framework (e.g., behavior extraction) are general, non-specialized tasks \cite{bhusal2024secure,fieblinger2024actionable} that are well within the capabilities of most LLMs, while larger models typically perform better.

\noindent\textbf{Hallucination Risk.}
The hallucinations risk of LLMs usage in our framework is partially limited by the closed, understanding-oriented nature of our tasks and the restricted graph scales during data synthesis. Future work will investigate this risk and potential mitigation strategies, such as consistency checking \cite{chen2024inside}.

\noindent\textbf{Limitations.}
APT hunting evaluation requires datasets with paired audit data and attack reports, which, to our knowled
ge, only DARPA datasets can satisfy. Other datasets that either lack complete audit records (e.g., StreamSpot \cite{manzoor2016fast}) or do not provide detailed attack reports (e.g., LANL \cite{turcotte2019unified}) are not suitable for this evaluation. Although our evaluation covers more than ten distinct attack types across different platforms, the scope of benchmarks remains limited. We leave the construction of more comprehensive and diverse APT datasets, encompassing broader attack behaviors and operating environments, as an important direction for future work.

\section{Conclusion}
We propose \textsc{APT-CGLP}, an end-to-end APT hunting system that enables cross-modal semantic matching between provenance graphs and CTI reports.
To mitigate data scarcity, we introduce a Graph2CTI module that employs LLM-driven heuristics to generate informative training pairs.
To enhance the usability of real-world CTI reports, the CoT reasoning capacity of LLMs is employed to extract actionable insights. Moreover, we design a multi-objective training framework that integrates contrastive learning and masked modeling to achieve multi-scale cross-modal semantic alignment.
Extensive experiments demonstrate that \textsc{APT-CGLP} delivers state-of-the-art threat hunting performance with minimal manual effort, low overhead, and high scalability.
%To tackle the challenge of limited training samples, we introduce a Graph2CTI module combined with a heuristic data augmentation approach to enrich the training data, thereby enhancing the generalization capability of \textsc{APT-CGLP} in detecting real-world APT activities.
%To mitigate the excessive noise in real-world CTI reports, \textsc{APT-CGLP} utilizes the Chain-of-Thought reasoning techniques of LLMs to extract critical attack knowledge to improve the usability of CTI reports.
%To bridge the substantial modality gap between provenance graphs and CTI reports, APT-CGLP employs a multi-objective training algorithm that integrates contrastive learning with masked data modeling, achieving precise semantic alignment at both coarse- and fine-grained levels.
%Comprehensive experiments demonstrate that \textsc{APT-CGLP} achieves threat hunting performance on par with state-of-the-art methods that rely on knowledge extraction with manual corrections, while maintaining low system overhead.
\section*{Acknowledgements}
This work has been partially supported by the National Natural Science Foundation of China (Grant Nos. U22B2028, 62372410), the Key Research Program of Hangzhou (Grant No. 2025SZD1A56), the Zhejiang Provincial Natural Science Foundation of China (Grant No. LD22F020002), the Key Research Program of Shaoxing (Grant No. 2025B11004), the Zhejiang Provincial Natural Science Foundation of China (Grant No. LZ23F020011), and the Zhejiang Province Leading Goose Program (Grant No. 2025C01013).

%\newpage
\bibliographystyle{ACM-Reference-Format}
\bibliography{ref}

\appendix
\section*{Appendix}
\appendix
\section{Activity Subgraph Sampling}\label{sampling_algorithm}
We design a subgraph sampling algorithm to extract semantic-rich subgraphs from provenance graphs, serving as templates for generating paired CTI reports to build a cross-modal learning dataset. 
%To ensure that the sampled subgraphs reflect realistic attack behaviors, we further incorporate several sampling constraints that guide the process and enhance the effectiveness of APT-CGLP.
As shown in Algorithm~\ref{subgraph sampling algorithm}, it adopts a three-layer Breadth-First Search (BFS) strategy and takes as input a provenance graph $pg$ and a size range $[minN, maxN]$, and outputs a set of subgraphs $SG_s$.

To better approximate real attack scenarios, we design a progressive sampling process.
In Layer 1 (lines 6-7), the algorithm samples process-type neighbors from socket nodes to capture initial attack entry points.
In Layer 2 (lines 8-10), it performs balanced sampling across different node types through BFS expansion, ensuring structural diversity in the subgraph.
In Layer 3 (lines 11-14), the subgraph is conditionally filled when its size falls below the minimum threshold, with additional nodes randomly sampled to reach a target size within the specified range.
Finally, each sampled subgraph is validated to ensure it contains all node types and falls within the predefined size range (lines 15-16), guaranteeing representativeness of genuine attack patterns.

\SetKw{Continue}{continue}
\begin{algorithm}[h!]
  \small 
  \caption{Activity Subgraph Sampling Algorithm}
  \label{subgraph sampling algorithm}
  \KwIn{Provenance graph $pg$; Size range $[minN, maxN]$}
  \KwOut{Subgraphs $SG_s$}
  \SetAlgoLined
  $SG_s \gets []$ \\
  $nei\_map \gets PrecomputeNeighbors(pg)$ \\
  $seed\_nodes \gets FilterByType(pg, "socket")$ \\

  \ForEach{$v \in seed\_nodes$}{
    $sg \gets \{v\}$ \\
    \tcp{Layer 1: Sample process neighbors}
    $N_1 \gets SampleByType(nei\_map[v], "process")$ \\
    $sg.update(N_1)$ \\

    \tcp{Layer 2: Balanced sampling across node types}
    $N_2 \gets BFS(N_1, nei\_map, sg)$ \\
    $N_2^{sampled} \gets TypeBalancedSample(N_2)$ \\
    $sg.update(N_2^{sampled})$ \\

    \tcp{Layer 3: Conditionally fill subgraph}
    \uIf{$|sg| < minN$}{
      $N_3 \gets BFS(N_2^{sampled}, nei\_map, sg)$ \\
      $N_3^{sampled} \gets RandomSample(N_3, minN - |sg|, maxN - |sg|)$ \\
      $sg.update(N_3^{sampled})$ \\
    }

    \tcp{Validate subgraph}
    \uIf{$minN \leq |sg| \leq maxN$ \& $CheckNodeTypes(sg)$}{
      $SG_s.add(sg)$ \\
    }
  }

  \Return{$SG_s$} \\
\end{algorithm}

\section{CTI Generation Prompt}\label{cti_generation_prompt}
\begin{figure*}[]
  \centering
  \includegraphics[width=7in]{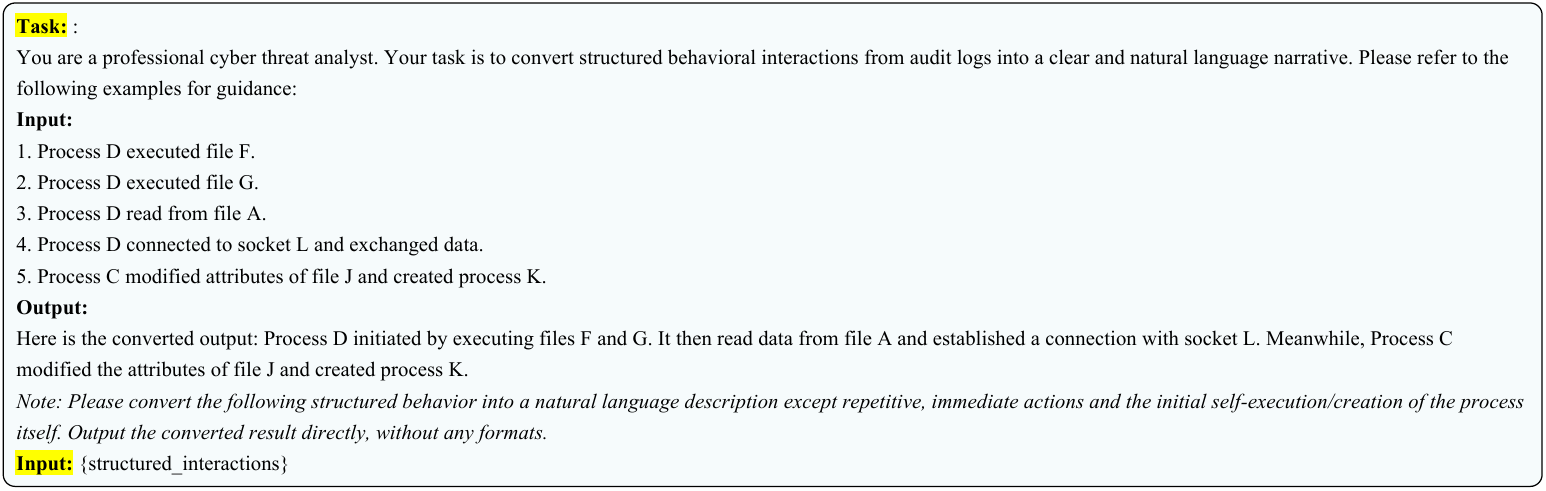}
  \caption{In-Context learning prompt for report generation.}
  \label{icl}
\end{figure*}

We harness LLaMA 's in-context learning capabilities to construct synthetic training data, thereby enriching cross-modal training datasets. 
This approach facilitates the generation of desired CTI-like reports by offering specific demonstrations within the prompt. We customize 5 various in-context learning demonstrations to guide the LLM in generating sufficient diverse synthetic reports, enabling the model to acquire robustness and generalization capabilities during training. A prompt example is illustrated in Figure \ref{icl}.

\section{CTI Denoising Prompt}\label{cti_clean_instructions}
\begin{figure*}[]
  \centering
  \includegraphics[width=7in]{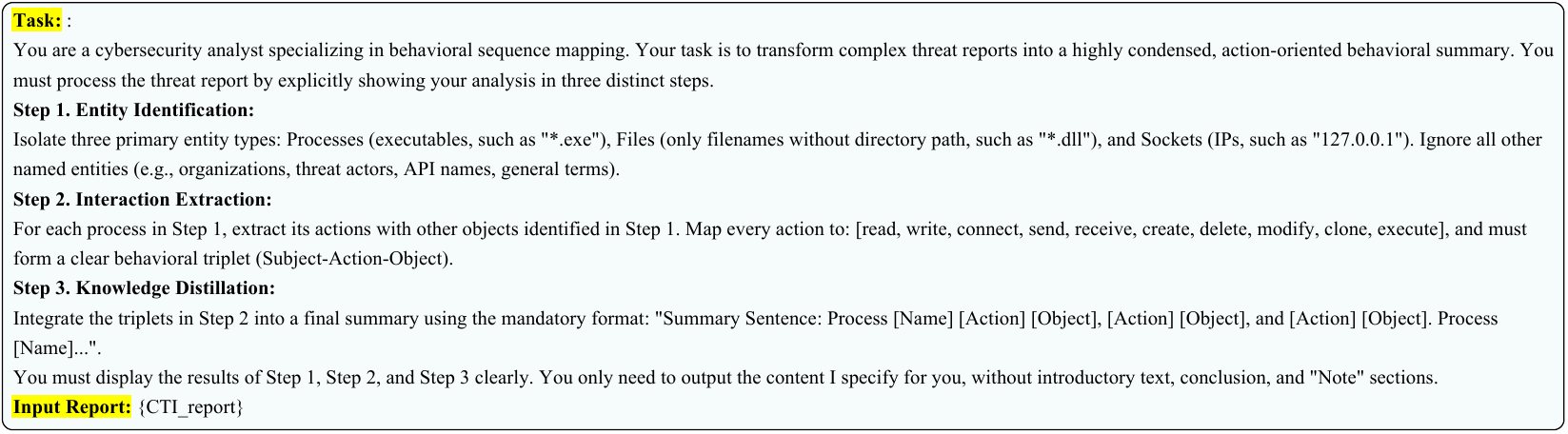}
  \caption{Chain-of-Thought prompt for CTI denoising.}
  \label{cot}
\end{figure*}

We employ LLaMA with Chain-of-Thought prompting to extract attack-relevant information from unstructured CTI reports.
This method reduces noise and produces coherent, actionable narratives suitable for downstream threat analysis. 
A prompt example illustrated in Figure \ref{cot} comprises three main parts: task description, multi-step processing instructions, and input data. 
%A case of CTI report denoising is shown in Figure \ref{cti_denoise_case}.

\begin{comment}
The prompt illustrated in Figure \ref{cot} comprises three main components: task description, multi-step processing instructions, and input data. 
The task description prompt provides LLaMA with crucial contextual information to improve comprehension and accuracy in processing CTI reports.
The multi-step instructions decompose the overall objective into a series of simple sub-tasks.
The model first identifies key system entities mentioned in the report, such as processes, files, and sockets.
It then extracts the interactions among these entities and maps them to a constrained set of interaction types (e.g., read, write, execute), ensuring alignment with standard system audit semantics.
Finally, the extracted behaviors are organized into a coherent sequence that highlights critical actions and preserves logical and temporal structure.
The output is a refined CTI report that faithfully captures essential attack behaviors, providing concise and reliable narratives to support threat analysis.
\end{comment}

\section{Graph Isomorphism Network} \label{app:graph_encoder}
We employ the Graph Isomorphism Network (GIN) \cite{gin} as graph encoder for its strong expressive power. 
%GIN achieves the comparable discriminative capability as the Weisfeiler-Lehman graph isomorphism test in distinguishing different graph structures, making it ideal for capturing complex semantics in provenance graphs.
The encoding process comprises aggregation, combination, and update phases, which iteratively refine node features by integrating semantic signals from their neighbors.
We initialize node features \{$h_v^{(0)} | v \in \mathcal{V}$ \} using a text encoder applied to node types and names, while edge features \{$e_{uv} | {u \in \mathcal{N}(v)}$\} are derived from their interaction types. 
At layer $k$, each node $v$ aggregates information from its neighbors through a learnable message function $g(\cdot)$: 
\[
\text{m}_v^{(k)} = \sum_{u \in \mathcal{N}(v)} g(h_u^{(k-1)}, e_{uv}).
\]

A trainable coefficient \(\epsilon^{(k)}\) then controls the relative contribution of the node's previous representation and the aggregated neighborhood message: 
\[
\text{s}_v^{(k)} = (1 + \epsilon^{(k)}) \cdot h_v^{(k-1)} + \text{agg}_v^{(k)}.
\]

The updated node representation is produced by applying a layer-specific MLP:
\[
h_v^{(k)} = \text{MLP}^{(k)}(\text{s}_v^{(k)}).
\]

After $k$ iterations, each node embedding encodes information from its $k$-hop neighborhood, effectively capturing local structural and semantic context. 
%Finally, a global summation READOUT aggregates node embeddings into a holistic graph-level representation.
%This approach enhances the expressive capability of graph structures while maintaining a balance between computational complexity and precision.

\begin{comment}
\begin{figure*}[]
  \centering
  \includegraphics[width=5.2in]{figs/cti_denoise_case.pdf}
  \caption{A case of CTI denoising, where subfigure (a) depicts a raw CTI report sourced from the web with line breaks removed, and subfigure (b) presents the CTI report after processing through the CTI denoising module.}
  \label{cti_denoise_case}
\end{figure*}
\end{comment}

\end{document}